\documentclass[12pt,preprint]{aastex}

\newcommand{\ie}{{\it{i.e.}}~}
\newcommand{\eg}{{\it{e.g.}}~}

\newcommand{\ha}{H$\alpha$~}

\newcommand{\lha}{L$_{H\alpha}$}
\newcommand{\pccm}{pc cm$^{-6}$}

\newcommand{\HI}{{\rm H\,\scriptstyle I}}
\newcommand{\HII}{{\rm H\,\scriptstyle II}}
\newcommand{\OIII}{{\rm [O\,{\scriptstyle III}]}}

\newcommand{\NII}{{\rm [N\,{\scriptstyle II}]}}

\newcommand{\HeI}{{\rm He\,\scriptstyle I}}

\shorttitle{FUV and \ha Imaging of DIG in Nearby Galaxies}
\shortauthors{Hoopes, Walterbos, \& Bothun}

\begin{document}


\title{Far Ultraviolet and \ha Imaging of Nearby Spirals: The OB Stellar Population in the Diffuse Ionized Gas}

\author{Charles G. Hoopes\altaffilmark{1}, Ren\'e A. M. Walterbos\altaffilmark{2}$^{,}$\altaffilmark{3}, \& Gregory D. Bothun\altaffilmark{4}}
\altaffiltext{1}{Department of Physics and Astronomy, Johns Hopkins University, 3400 North Charles Street, Baltimore, MD 21218; choopes@pha.jhu.edu}
\altaffiltext{2}{Department of Astronomy, New Mexico State University, MSC 4500, Box 30001, Las Cruces, New Mexico 88003; rwalterb@nmsu.edu}
\altaffiltext{3}{Visiting Astronomer, Kitt Peak National Observatory
and Cerro-Tololo Inter-American Observatory, National Optical Astronomy
Observatories, operated by the Association of Universities for
Research in Astronomy, Inc. (AURA) under cooperative agreement with
the National Science Foundation.}
\altaffiltext{4}{Physics Department, University of Oregon, Eugene, OR 97403; nuts@bigmoo.uoregon.edu}

\begin{abstract}
We have compared \ha and far ultraviolet (FUV) images of 10 nearby
spirals, with the goal of understanding the contribution of field OB
stars to the ionization of the diffuse ionized gas (DIG) in spiral
galaxies. The FUV images were obtained by the Ultraviolet Imaging
Telescope (UIT) and the \ha images were obtained using various
ground-based telescopes. In all of the galaxies, the
F$_{H\alpha}$/F$_{UIT}$ flux ratio is lower in the DIG than in the
\ion{H}{2} regions. This is likely an indication that the mean spectral
type for OB stars in the field is later than that in \ion{H}{2}
regions. Comparison of the $N_{Lyc}/L_{UIT}$ ratio with models of
evolving stellar populations shows that the stellar population in the
DIG is consistent with either an older single burst population or a
steady state model with constant star formation and an initial mass
function (IMF) slope steeper than $\alpha$=2.35. The steady state
model is probably a more realistic representation of the stellar
population outside of \ion{H}{2} regions. The steep IMF slope
simulates the steep present-day mass function slope expected for field
OB stars, and does not necessarily indicate that IMF slope is actually
steeper than $\alpha$=2.35. We compared the F$_{H\alpha}$/F$_{UIT}$
ratio in the DIG of these galaxies with that in M33, in which the
field OB stellar population has previously been investigated using
Hubble Space Telescope images. If the mean spectral types of stars in
\ion{H}{2} regions and in the DIG are the same as in M33, and the difference
in extinction between DIG and \ion{H}{2} regions is constant among galaxies,
then the analysis suggests that field stars are important sources of
ionization in most galaxies, and may be the dominant source in some
galaxies. The F$_{H\alpha}$/F$_{UIT}$ ratio is correlated with \ha
surface brightness in both DIG and \ion{H}{2} regions, although there
is a large scatter in faint \ion{H}{2} regions, which may be due to
undersampling the IMF in regions with a low total mass of stars
formed. The F$_{H\alpha}$/F$_{UIT}$ ratio is often highest in the
centers of galaxies and in the spiral arms, which is also where the
DIG is brightest. This can be explained if the extinction is greater
in these regions, or if the fraction of DIG ionized by leakage is
lower in the inter-arm regions.

\end{abstract}



\section{Introduction}

The details of the relationship between massive stars and the ISM are
crucial to understanding how galaxies evolve. One place where this
interaction is critical, and not well understood, is in the diffuse
ionized gas (DIG). This is the dominant component of ionized gas in
terms of mass, contributing 95\% of the ionized gas mass in the Milky
Way \citep{re91}, and requiring the equivalent of 30 to 50\% of the
ionizing photons of OB stars to remain ionized in almost all of the
external spirals studied \citep{wb96}. The DIG is clearly an important
physical manifestation of energy transfer from stars to gas, yet the
mechanisms through which this transfer occurs remain unclear. The
large energy requirement, the morphological association of DIG with
\ion{H}{2} regions \citep{wb94,f96,hwg96,g98,gwth98}, and the spectral
properties of the DIG \citep{gwb97,whl97,gwb99,hrt99} all very
strongly suggest OB stars as the ionization source, although there are
indications that other sources, such as old supernova remnants
\citep{smh00} or turbulent mixing layers \citep{ssb93} also play a
role \citep{whl97}. However, problems transporting the ionizing
photons from OB stars to the DIG remain unsolved. Specifically, if the
ionizing stars are in \ion{H}{2} regions, are there enough density-bounded
\ion{H}{2} regions leaking ionizing photons, as well as paths through the
$\HI$ in spiral disks, for these photons to reach the DIG?

An alternative is that field OB stars ionize some of the DIG, reducing
the amount of leakage from \ion{H}{2} regions required. Field OB stars
have been studied in the Milky Way \citep{tlp74}, the Magellanic
Clouds \citep{mas95b}, and M33 \citep{pw95,hw00}. There are several
possible explanations for the presence of these OB stars outside of
\ion{H}{2} regions. They may form in \ion{H}{2} regions, but then diffuse out
of the \ion{H}{2} region into the diffuse gas, or they may survive as an
\ion{H}{2} region is destroyed by supernovae (SNe) from other massive
stars. Both of these processes would produce a population in the field
that is weighted toward later-type ionizing stars. A late-type (B0-O9)
ionizing star has a longer lifetime than an earlier-type star,
increasing the probability that it will live long enough to move an
appreciable distance from its birthplace, or survive as the \ion{H}{2}
region is destroyed when the more massive stars explode in
SNe. Another possibility is that field OB stars actually form in the
field, as suggested by \cite{mas95b} for the Magellanic Clouds. Those
authors defined their sample of field stars so that it would not
include stars which might have traveled out of an \ion{H}{2}
region. However, they also found that the initial mass function (IMF)
for field stars has a steeper slope than in \ion{H}{2} regions, making the
very early-type OB stars rare. All of these mechanisms, then, lead to
a field population dominated by later-type OB stars. These stars must
ionize some fraction of the DIG, but since they are predominantly
later-types, a large number is needed to explain all of the DIG. The
presence of just one O3 star in the field is equivalent to 50 B0
stars.

A direct way to determine the fraction of DIG that is ionized by field
OB stars is to look for OB stars in and out of \ion{H}{2} regions,
determine their spectral types and ionizing luminosities (N$_{Lyc}$),
and compare this to the energy required to maintain the observed \ha
luminosity. This was done for \ion{H}{2} regions in the LMC, and several
were found that may be leaky \citep{ok97}. \cite{hw00} studied OB
stars in \ion{H}{2} regions and DIG in M33, using {\it Hubble Space
Telescope} (HST) far-ultraviolet (FUV) and optical images. The FUV
information is crucial for assigning ionizing luminosities to massive
stars, as there is a severe color degeneracy for hot stars at optical
wavelengths \citep{mas85,mas95a}. Even in the FUV the color degeneracy
is a problem, but when combined with the FUV and optical absolute
magnitudes, the problem is reduced to a level where values of
N$_{Lyc}$ good to the equivalent of about $\pm1$ spectral subtype can
be obtained.

The current paper builds on the results of \cite{hw00}, so we briefly
describe the analysis carried out in that paper. The goal was to
determine the ionizing luminosities of stars in \ion{H}{2} regions and in
the field, in order to compare with ionizing luminosity inferred from
the \ha emission in each environment.  Magnitudes were measured from
the WFPC2 images in four filters, F170W, F336W, F439W, and F555W,
along with the statistical uncertainty and flat fielding uncertainty,
for each star. Color magnitude diagrams (CMDs) for each \ion{H}{2} region
and DIG region were plotted and compared to theoretical CMDs
\citep{s93}. The extinction correction required to place the observed
main sequence on the theoretical main sequence was then determined,
resulting in a mean extinction correction for each region. A
conservative estimate of the uncertainty in determining this
correction was combined with the previously measured uncertainty in
the photometry.

The extinction corrected magnitudes in four bands for all of the stars
were compared with the predicted magnitudes for massive stars from the
CoStar models \citep{sd97}, and N$_{Lyc}$ from the best matching model
was assigned to each star. The uncertainty in N$_{Lyc}$ was determined
by finding the models with the largest and smallest ionizing
luminosities that matched within the 1$\sigma$ uncertainty in the
magnitudes (which included statistical, flat-field, and extinction
uncertainties). The N$_{Lyc}$ values for all stars in a given
\ion{H}{2} region or DIG region were summed together, and compared
with N$_{Lyc}$ required by the \ha emission.  In the regions of M33
that were investigated, OB stars outside of \ion{H}{2} regions can ionize an
appreciable amount of the DIG, $40\%\pm12\%$, while the OB stars in \ion{H}{2}
regions account for $107\%\pm26\%$ of the observed H$\alpha$ luminosity of
the \ion{H}{2} regions. The good agreement between the mean predictions and
observations for \ion{H}{2} regions gives confidence that the DIG result is
not affected by systematic uncertainties. Thus, even though estimates
of ionizing luminosities for individual stars have substantial
uncertainty since spectral classification from photometry alone is not
very accurate, the overall result seems robust. The sizable
contribution of field OB stars to the ionization of the DIG relaxes
somewhat the need for leaky \ion{H}{2} regions, and emphasizes the fact that
OB stars outside of \ion{H}{2} regions have a large impact on the ISM.

In this paper we investigate whether the source of the DIG ionization
in M33 is similar to other spirals.  In more distant spirals HST
cannot resolve individual stars.  We have used images acquired by the
{\it Ultraviolet Imaging Telescope} (UIT) to yield global FUV
information, which can then be compared to M33, in which we have both
HST and UIT data.  The resolution of the UIT images is approximately
2$^{\prime\prime}$ or $\sim$80 parsecs at the typical distances of our
targets.  Thus we can well resolve the FUV field.  Comparing the
spatial distribution of FUV emission to that of the \ion{H}{2} regions
(from narrow band \ha imaging) in our target galaxies will allow for
an assessment of the sources of ionization in a manner similar to that
done by \cite{hw00}.

\section{Data Reduction and Analysis}

The sample of galaxies for which we analyzed \ha and FUV images is
described in Table 1. Ten galaxies were included, including M33 which
was studied previously \citep{hw00}. The sample spans a range of
Hubble types, but early-type spirals are under represented. Most of
the galaxies are nearby, although NGC 1512 and NGC 1566 are
beyond 15 Mpc. A metallicity estimate for each galaxy was taken from
\cite{zkh94}, using their value of 12+log(O/H) at a radius of 3 kpc
from the nucleus.

Table 2 summarizes the \ha data for each galaxy. The reduction for M33
was described in \cite{g98} and \cite{hw00}, M81 and M51 were
discussed in \cite{gwth98}, and M101, NGC 628, NGC 925, and NGC 2903
were discussed in \cite{g98}. Both line and nearby continuum images
were obtained, and the continuum image was scaled to the line image by
comparing the fluxes of foreground stars, and then subtracted to
produce a line-only image. The calibration method for each galaxy is
given in Table 2. Most were calibrated using observations of a
standard star, but a few were calibrated using the R-band magnitude of
the galaxy and the shape of the narrow band filter, and a few were
calibrated by comparing the luminosities of specific \ion{H}{2} regions to
published values. Three galaxies, NGC 3031 (M81), NGC 5194 (M51) and
NGC 5457 (M101) were observed through \ha filters wide enough to
include a contribution from $\NII$ $\lambda6548+6584$ \AA. In these
cases, 20 to 30\% of the luminosity my arise from $\NII$. We have not
corrected for this contamination, but it will be addressed in the
relevant sections.

Table 3 summarizes the UIT data for the sample. Details of the UIT
instrument and data characteristics can be found in \cite{s97}. The
UIT instrument was flown on two occasions: the Astro-1 mission aboard
the space shuttle Columbia in December, 1990, and the Astro-2 mission
aboard the space shuttle Endeavor in March, 1995. The galaxies were
observed by the FUV ($\sim$ 1500\AA) camera, through one of two
filters: B1 (1520\AA, $\Delta\lambda$ 354\AA) and B5 (1615\AA,
$\Delta\lambda$ 225\AA). The images were originally produced on film,
and later digitized on the ground. The images were calibrated by
observing standards stars observed previously with the International
Ultraviolet Explorer, and are estimated to be accurate to 15\%
\citep{s97}. Nonlinearity at low flux levels introduces further
uncertainty. The effects of nonlinearity have been found to be as high
as 20\%, and the effects vary significantly between exposures
\citep{bk01}. We adopt 15\% uncertainty for \ion{H}{2} region UIT flux, and
25\% (15\% + 20\% in quadrature) for the fainter DIG UIT flux. The UIT
images have an angular resolution of about 2$^{\prime\prime}$.

For each galaxy the DIG was isolated in the \ha image using the
masking procedure described in \cite{hwg96}. The \ha image was
smoothed with a median filter using a linear length scale of 1
kpc. The smoothed version was subtracted from the original, leaving an
image of the small scale structure only, the smoothly varying
component having been subtracted away. On this image a mask was
created by replacing pixels with values less than 50
\pccm~in emission measure with one, and pixels with values greater
than 50 \pccm~with zero. The resulting mask thus has a value of zero
in the bright, small scale structures (\ie \ion{H}{2} regions), and one
everywhere else.  The original (unsmoothed) \ha image was then
multiplied by the mask image, leaving an image of only the DIG, with
zeros over the \ion{H}{2} regions. This is called the DIG \ha image. A DIG
FUV image was also created by multiplying the FUV image with the same
mask (the mask defined on the \ha image). An \ion{H}{2} region mask was
created by inverting the DIG mask, and this was multiplied with the
\ha and FUV images, producing an \ion{H}{2} region \ha image and an
\ion{H}{2} region FUV image.

Using the \ion{H}{2} region \ha image, circular apertures were defined by
eye for as many \ion{H}{2} regions as possible. The fluxes of these
regions were then measured on the original (unmasked) \ha and FUV
images. The \ion{H}{2} region image is used only to find the \ion{H}{2}
regions with no confusion from the DIG. A local background was defined
for each \ion{H}{2} region as the mode of the pixel values in an annulus
around the aperture. This background was subtracted from each pixel in
the aperture.

The DIG fluxes were measured in square 500$\times$500 pc apertures
which covered the galaxy. The fluxes were measured on the DIG images
with the \ion{H}{2} regions masked out. A lower cutoff equal to the
1-$\sigma$ noise level in the \ha image was applied to the resulting
images, in order to remove apertures which did not include any
DIG. For each galaxy there were a small number of apertures that
contained \ha flux but no FUV flux (both DIG and \ion{H}{2} regions after
background subtraction). These were not included in the final
analysis. These procedures follow exactly those used to analyze M33 in
\cite{hw00}.

\section{Results}

\subsection{The H$\alpha$/FUV Ratio}

In Table 4 we list the observed \ha and FUV properties of the sample
galaxies. The global F$_{H\alpha}$/F$_{UIT}$ DIG and \ion{H}{2} region
ratios were measured using the masked images.  The diffuse fraction is
the ratio of the DIG \ha luminosity (measured on the masked DIG image)
to the total \ha luminosity of the galaxy. The FUV diffuse fraction is
the fraction of the total FUV luminosity that arises in the DIG. The
uncertainty in \lha\ and in the \ha diffuse fraction was found by
varying the continuum scale factor by $\pm$3\%. The uncertainty in the
flux ratios is the combination of the uncertainty in \lha\ and the
uncertainty in L$_{UIT}$, which is the 15\% calibration uncertainty in
HII regions, and 25\% (15\% calibration and 20\% nonlinearity) in DIG
\citep{s97,bk01}. For the ratio of total luminosities we
conservatively use the 25\% uncertainty for L$_{UIT}$. In all cases
except NGC 1512, the \ha diffuse fraction is roughly between 30 and
50\%.  The H$\alpha$/FUV ratios are higher in the DIG than in \ion{H}{2}
regions. This can also be seen in the FUV diffuse fraction, which is
much higher than the \ha diffuse fraction for all of the galaxies. In
the FUV, a greater fraction of light is found in the field than in
H$\alpha$, due to the fact that the FUV band includes light from
later-type B and A stars. In NGC 1512, 39\% of the
\ha flux is emitted by the nuclear star-forming ring. If this region
is excluded, the \ha diffuse fraction rises to about 20\%.

Figure 1 shows the histograms of the F$_{H\alpha}$/F$_{UIT}$ flux
ratio for the galaxies observed in the B1 filter, and Figure 2 shows
the galaxies observed through the B5 filter. The histograms show the
observed values; no correction for extinction has been made at this
point. M33 is shown twice because it was observed through both the B1
and B5 filters. The three galaxies with $\NII$ contamination are not
noticeably different, but the correction to the ratio would be only
about 0.1 in the logarithmic plots.

In all of the galaxies in the sample, \ion{H}{2} regions have a higher
F$_{H\alpha}$/F$_{UIT}$ ratio than DIG regions. \cite{hw00}
interpreted this trend seen in M33 as a hotter stellar population in
the \ion{H}{2} regions, and confirmed the interpretation through HST WFPC2
imaging of individual OB stars in \ion{H}{2} regions and DIG. The \ha
emission results from ionizing photons, so the F$_{H\alpha}$/F$_{UIT}$
ratio is really a ratio of ionizing/non-ionizing UV photons. The
spectra of hotter stars will have a higher ratio of ionizing to
non-ionizing photons (\ie a {\it bluer} spectrum) than that of cooler
stars. The difference suggests that the population of ionizing stars
in the field is dominated by cooler, later-type OB stars, in
qualitative agreement with the mechanisms for forming field stars
discussed in section 1.

\subsection{Comparison with Starburst99 Models}

In this section we compare the F$_{H\alpha}$/F$_{UIT}$ ratios to
models of evolving stellar populations, to see whether the observed
ratios can be reproduced by a reasonable population of stars.  Figures
3 and 4 show the same histograms as Figures 1 and 2, but now they are
shown sideways for comparison with the models. We have also converted
the H$\alpha$ luminosity to the number of ionizing photons $N_{Lyc}$,
assuming case B recombination \citep{o89} and neglecting absorption of
ionizing photons by dust. The histograms now include a rough
extinction correction, which depends upon the models also shown in the
figures. First we will describe the models, and then we will describe
the correction for extinction.

The Starburst99 stellar population evolution model \citep{l99} was
used to simulate the conditions in \ion{H}{2} regions and in the
field. This model simulates the evolution of a star cluster through
time, and allows one to change parameters, such as the initial mass
function (IMF), metallicity, and star formation rate. We ran two types
of models. One set has a single burst of star formation, which
represent an \ion{H}{2} region. The other is a steady state model with
constant star formation, which roughly approximates the conditions in
the field, in which a relatively small population of young stars is
superimposed on a much larger population of old stars.

The burst models begin with a high $N_{Lyc}/L_{UIT}$ ratio, as the
cluster is dominated by hot young stars with large ionizing
luminosities. As the cluster ages and the most massive stars evolve
and die, $N_{Lyc}$ drops, while $L_{UIT}$ stays relatively constant
since it contains emission from later-type B stars as well as O stars,
so the $N_{Lyc}/L_{UIT}$ ratio falls. Thus the $N_{Lyc}/L_{UIT}$ ratio
is a good indicator of the age of a burst population
\citep{hill95,hill98,s00}. The steady state models reach an
equilibrium value.

The metallicity of each type of model was varied to include LMC like
abundances(z=0.08), solar (z=0.20) and twice solar (z=0.40). The
metallicity chosen for each galaxy was based on the measurements of
\cite{zkh94} in Table 1. We also varied the IMF slope to
include $\alpha$=2.35, 3.00. and 3.50. The purpose was to simulate a
steeper mass function slope, which might arise from the processes
which that produce field OB stars (see section 1). The steeper mass
function results in a lower $N_{Lyc}/L_{UIT}$ ratio, as the most
massive O stars become rare compared to the number of less massive OBA
stars. For \ion{H}{2} regions we show only the standard $\alpha$=2.35 IMF
models, as there has been no indication of variation of the IMF in
\ion{H}{2} regions \citep{mas98}.

In order to make a rough estimate of the extinction in each galaxy, we
assume that the single burst models are appropriate for the brightest
\ion{H}{2} regions in each galaxy (\eg von Hippel \& Bothun 1990). The
brightest regions probably have the largest number of O stars, and
thus the IMF is more likely to be well sampled. Under-sampling the IMF
can have a large effect on the $N_{Lyc}/L_{UIT}$ ratio, and this is
likely to happen if the number of O stars is low. An extreme case is a
single O3 star and no other UV emitting stars, resulting in a very
high $N_{Lyc}/L_{UIT}$ ratio. This is why we do not use the \ion{H}{2}
regions with the highest observed $N_{Lyc}/L_{UIT}$ to define the
extinction correction, as one might be tempted to do since the models
predict that these should be the youngest \ion{H}{2} regions.

For each galaxy we found the average of log $N_{Lyc}/L_{UIT}$ for the
ten brightest \ion{H}{2} regions (only 3 for M33, which lacks very bright
\ion{H}{2} regions). The average value was compared to the prediction of
the single burst model (for the metallicity of that galaxy) at 1
Myr. We chose this age based on the assumption that part of the reason
these \ion{H}{2} regions are bright is that they are young, since the
models in Figures 3 and 4 show that after a few Myr the ionizing
luminosity of a star cluster drops drastically as the most massive
stars die. The difference between the mean log $N_{Lyc}/L_{UIT}$ and
the model value is the extinction correction for that galaxy. The
histograms in Figures 3 and 4 have been corrected by that
amount. Table 5 lists the shifts used for each galaxy, and the
corresponding E(B-V) for two extinction laws. Since the magnitude of
the correction was measured, the particular extinction law does not
matter, but the values are shown for reference. Published values of
E(B-V) for these galaxies are also listed in Table 5.

Correcting for extinction moves the histograms down in the plots,
because the correction for $L_{UIT}$ is larger than that for $N_{Lyc}$
(measured through H$\alpha$). We do not specifically correct the
ionizing photon flux for absorption by dust, however making the \ion{H}{2}
regions equal to the model predictions should also correct for
this. \cite{mw97} find that about 25\% of the ionizing photons emitted
by stars in \ion{H}{2} regions are absorbed by dust. The effect of
absorption by dust on the observed histograms is to move them down, by
making the $N_{Lyc}/L_{UIT}$ lower, so the correction for this works
in the opposite sense of correcting the \ha and FUV light for
extinction. This may be part of the reason that the extinction values
in Table 5 are lower than the previously determined values for those
galaxies.

This is an average correction for all \ion{H}{2} regions (and DIG). Part
of the scatter in the histogram is probably due to differences in
extinction from region to region, but we have no way to correct for
this. A large assumption in this process is that the extinction in the
DIG is the same as that for \ion{H}{2} regions. In reality it is likely
that the extinction is somewhat different due to several processes, an
example being the destruction of large grains due to shocks in \ion{H}{2}
regions \citep{jth96}. In both M31 \citep{gwb97} and M33 \citep{hw00}
the difference has been small, about $\Delta$A$_V$=0.3. This much
difference would move the DIG histograms in Figures 3 and 4 up by 0.22
using the Milky Way extinction law, and 0.32 using the LMC extinction
law. It is also possible that the fraction of $N_{Lyc}$ absorbed by
dust is different in the DIG. We will discuss the effects of a
difference in extinction where it is relevant.

The extinction corrections for NGC 1313, NGC 1512, NGC 1566, and NGC
925, are particularly low. These are also the four galaxies (besides
M33) that were observed through the B5 filter. The extinction
correction derived for the two M33 observations are remarkably
consistent, so there does not seem to be a problem with the B5
filter. There may be a calibration problem with the FUV images of
these galaxies, or there may be a real difference between these and
the other galaxies, such as a higher fraction of $N_{Lyc}$ absorbed by
dust. In any case, the results for these galaxies should be treated
with caution.

After the extinction correction, most of the galaxies look very
similar to M33. The DIG ratios are consistent with a steady state
model, or with a single burst model older than that for \ion{H}{2}
regions. The steady state model seems like a more realistic model for
the field stellar population than an older burst, as the steady state
model represents a disk built up through star formation over several
Gyr, with a much smaller population of young stars. The DIG histograms
exhibit even better agreement with the steady state models with
somewhat steeper IMF slopes, which was also seen in M33. As already
stated, the steep IMF slopes in the models simulate a steep {\it
present-day} mass function slope, which may be a result of the way
\ion{H}{2} region stars become field stars, and not necessarily an
indication of how field stars actually form.

\subsection{Analysis of the Observed Ratios}

The comparison with the Starburst99 models supports the general
conclusion that the field OB star populations in these galaxies are
similar to that in M33, and that they are likely to ionize a similar
fraction of the DIG. However, the differences that do exist between
these galaxies may indicate real differences in the amount of leakage
or field star ionization present. A comparison of the mean
F$_{H\alpha}$/F$_{UIT}$ ratios may elucidate some of these differences
or lack thereof. Instead of comparing to models, we will compare the
observed F$_{H\alpha}$/F$_{UIT}$ ratios to those seen in M33, where we
have independent information on the fraction of DIG ionized by field
OB stars. In this section we use only the observed values, without the
extinction correction described in the last section.

The analysis is complicated by the important effects of extinction at
ultraviolet wavelengths, and the possible differences in stellar
populations in the field and in \ion{H}{2} regions due to varying star
formation histories, metallicity, or IMF slopes. To mitigate these
complicating factors, we will begin the analysis by making some
simplifying assumptions. First we will assume that the \ion{H}{2} region
stellar population is the same among galaxies. This means that the
mean intrinsic F$_{H\alpha}$/F$_{UIT}$ ratio in \ion{H}{2} regions for all
of the galaxies will be the same. The variation seen in the observed
ratio (Table 5) is within the range that can be explained by
differences in the amount of extinction, so we have no evidence that
this assumption is false.  Second, we assume that the relative
extinction in \ion{H}{2} regions and DIG is the same for all galaxies. By
relative, we mean that there may be more or less extinction in DIG
compared to \ion{H}{2} regions, but the difference between \ion{H}{2} region
and DIG extinction will be constant for all galaxies. For M31 and M33,
the only two galaxies where this has been measured, this assumption
appears to be true \citep{gwb97,hw00}.

With these assumptions we can then use the measured \ion{H}{2} region
ratio to anchor the DIG ratio, independent of extinction. The
difference between the \ion{H}{2} region ratio and the DIG ratio can then
be compared among galaxies. This approach does not suffer from the
effects of any errors that might have been made in the extinction
correction made using the starburst models. Table 6 lists for each
galaxy the differences in the logarithm of the mean observed
F$_{H\alpha}$/F$_{UIT}$ ratios for \ion{H}{2} and DIG. If the assumptions
we have made are true, then the difference in this value for different
galaxies is completely due to a difference in the
F$_{H\alpha}$/F$_{UIT}$ ratio in the DIG, since we have assumed that
the \ion{H}{2} ratio is constant, and the \ion{H}{2}/DIG relative comparison
removes the effects of extinction.

The interpretation of these values is still complicated. Since the
\ion{H}{2} regions are ionized by local stars with no leakage of ionizing
photons into the system, the simplest interpretation is to consider
them as the zero-leakage reference point for the DIG. In this case
differences between the \ion{H}{2} region and DIG ratios would be
interpreted as being due to leakage, with larger differences
indicating more leakage. However, all of the DIG
F$_{H\alpha}$/F$_{UIT}$ ratios are smaller than the \ion{H}{2} region
ratios. The largest differences arise from the smallest DIG ratios, so
taking this scenario to the extreme means that a ratio of zero (\ie no
\ha emission) corresponds to maximum leakage. The key point is that
leakage should always increase the F$_{H\alpha}$/F$_{UIT}$ ratio in
the DIG by adding \ha flux but no FUV flux. Since the DIG ratios are
always lower than the \ion{H}{2} region ratios, leakage cannot be the only
factor affecting the ratios.

The other factor is the stellar population in the field, which is
different from that in \ion{H}{2} regions. \cite{hw00} found that the mean
spectral type of stars in the field is later than that in \ion{H}{2}
regions, a conclusion also found for the LMC, SMC, and Milky Way
\citep{ mas95b}. Later type stars have lower ratios of L$_{Lyc}$/FUV
luminosity (\ie more FUV luminosity relative to their ionizing
luminosity). This results in lower F$_{H\alpha}$/F$_{UIT}$ ratio for
DIG than \ion{H}{2} regions in the absence of leakage. Because of this
fact, \ion{H}{2} regions do not represent the zero-leakage reference point
for DIG.

However, we can use the M33 ratio as a reference point for the DIG in
other galaxies, since we know that field OB stars ionize $\sim40$\% of the
DIG in that galaxy \citep{hw00}. Table 6 shows the difference in the
F$_{H\alpha}$/F$_{UIT}$ ratio relative to M33. Here we make another
assumption: the mean spectral type of the field stars, while different
from that in \ion{H}{2} regions, is the same in all of the galaxies. Then
the difference in the ratio can be understood as being due solely to a
difference in the fraction of DIG ionized by leakage. The
F$_{H\alpha}$/F$_{UIT}$ ratio for the DIG that is ionized only by
field stars remains the same, independent of the number of field
stars, since the mean spectral type is constant. Then any change in
the DIG F$_{H\alpha}$/F$_{UIT}$ ratio is caused by the addition or
subtraction of \ha flux from DIG ionized by leakage.

We illustrate this scenario using M51 as an example. The observed
difference in the mean log F$_{H\alpha}$/F$_{UIT}$ ratio for DIG and
\ion{H}{2} regions for M51 is 1.09, while for M33 it is 0.81. Since this
is a difference of logarithms it is actually a ratio, and if the
intrinsic \ion{H}{2} region F$_{H\alpha}$/F$_{UIT}$ ratios are the same
for both galaxies, this difference means that the M51
F$_{H\alpha}$/F$_{UIT}$ ratio for DIG is 0.52 that in M33. We assume
that this difference is totally due to a change in the fraction of DIG
ionized by leakage. In M33, 40\% of the DIG is ionized by field stars,
so 60\% of the observed \ha in the DIG come from leakage. The \ha flux
from leakage in M51 has to be reduced to get the lower observed
F$_{H\alpha}$/F$_{UIT}$ ratio, resulting in a fraction of DIG ionized
by leakage of 23\%, and 77\% ionized by field OB stars. The fraction
of DIG ionized by field stars for each galaxy based on these
assumptions is shown in Table 6. 

In this scenario, field OB stars dominate the ionization of the DIG in
M51 and NGC 628, NGC 1313 and NGC 1512 have smaller contributions from
field stars, and the other galaxies are similar to M33. The large
\ha luminosities of M51 and NGC 628 indicate that they have both
undergone recent galaxy-wide star formation events. If this is
connected with the dominance of the field OB stars, it suggests that
these events somehow enhance the creation of field OB stars, but do
not affect the amount of DIG ionized by leakage. This is highly
speculative however, since other galaxies in the sample have had
recent star formation, but do not appear to exhibit this effect. The
\ha filter used for M51 contains a contribution from $\NII$, and the
$\NII$/\ha ratio is higher in the DIG, which may affect the observed
difference. However, correcting for this would make the
F$_{H\alpha}$/F$_{UIT}$ ratio lower in the DIG compared to \ion{H}{2}
regions, and make the difference between \ion{H}{2} regions and DIG even
larger, raising the predicted contribution of field OB stars even
higher. 

If the assumption that the field star populations are similar in
spectral type in all galaxies is in error, the analysis becomes more
complicated. In this case the F$_{H\alpha}$/F$_{UIT}$ ratio for DIG
ionized by field stars will be different, higher in galaxies where the
mean spectral type is earlier, and lower if the mean spectral type is
later. Put another way, if the mean spectral type is earlier, then a
field star population with a given amount of FUV luminosity can ionize
more DIG (resulting in higher \ha luminosity) than in a galaxy with a
later mean spectral type in the field. The F$_{H\alpha}$/F$_{UIT}$
ratio for DIG ionized by such a population of field OB stars would be
closer to that in \ion{H}{2} regions. In M51, for example, the large
difference between the \ion{H}{2} regions and DIG ratios suggests the DIG
ratio is low, which in this scenario would imply a field OB star
population with a later mean spectral type than in M33. Disentangling
this effect from a change in the amount of leakage is not possible
with these data, as the two have the same effects on the
F$_{H\alpha}$/F$_{UIT}$ ratio.

\subsection{Radial and Surface Brightness Variations}

For the following analysis we selected four galaxies in order to make
the figures easier to read.  Figure 5 shows the
F$_{H\alpha}$/F$_{UIT}$ as a function of \ha surface brightness for
\ion{H}{2} regions and DIG for four of the galaxies with the best FUV
images: M33 (B1), M51, M81, and M101. The ratio is correlated with
surface brightness in both environments. In the \ion{H}{2} regions, a high
F$_{H\alpha}$/F$_{UIT}$ can arise from either a hotter (\ie younger)
stellar population, or higher extinction. Although there is a trend
for the F$_{H\alpha}$/F$_{UIT}$ to increase with \ha surface
brightness, the brightest \ion{H}{2} regions do not have the highest
F$_{H\alpha}$/F$_{UIT}$ ratios (this is most apparent in M101 or M81,
for example). In fact the scatter in the ratio tends to increase as
the \ha intensity decreases from the brightest to moderately bright
regions, so that the regions with the highest F$_{H\alpha}$/F$_{UIT}$
are only moderately bright regions, while the brightest regions have a
more moderate ratio (although less so for M33). This scatter is larger
than the uncertainty in the ratio, which even at a very conservative
30\% in the low luminosity regions is only about $\pm$0.1 in the
logarithmic plot. These high ratios may be caused by undersampling of
the IMF in regions with low total mass of stars formed (\eg Oey \&
Clarke 1998). If the undersampling causes the mass function to be
top-heavy (more very massive early O stars and few or no less massive
late O and B stars) a high F$_{H\alpha}$/F$_{UIT}$ ratio will
result. Our choice to use the brightest \ion{H}{2} regions to define the
extinction correction for the histograms in Figures 3 and 4 rather
than those with the highest F$_{H\alpha}$/F$_{UIT}$ ratio was designed
to avoid these stochastic effects.

Figure 6 shows the F$_{H\alpha}$/F$_{UIT}$ ratio as a function of
projected distance from the center of the galaxy. The higher ratios in
the \ion{H}{2} regions are clearly seen again, and several of the galaxies
show higher ratios in the DIG toward the center of the galaxy,
although it is not apparent in M101. In the total sample about half of
the galaxies exhibit this trend. Possibly related to this effect is
the fact that the DIG near the centers of spirals is usually brighter
than DIG further out (Greenawalt 1998).  The radial trend could be an
effect of extinction. Higher extinction in the centers of galaxies
would produce higher F$_{H\alpha}$/F$_{UIT}$ ratios. 

Figure 7 shows the F$_{H\alpha}$/F$_{UIT}$ ratio in DIG regions as a
function of projected distance from the nearest \ion{H}{2} region. In
general, it appears that at larger distances from \ion{H}{2} regions, the
DIG regions have lower F$_{H\alpha}$/F$_{UIT}$, while near \ion{H}{2}
regions, the DIG regions can have low or high ratios (\ie, the high
ratio DIG regions are missing far from \ion{H}{2} regions). A higher
F$_{H\alpha}$/F$_{UIT}$ in the DIG can either mean that the field OB
star population there is dominated by earlier-type stars, or it may
indicate a larger fraction of leakage of ionizing photons into the
region. Both scenarios seem possible near \ion{H}{2} regions, and we
cannot distinguish between the two with these data. We explore this
further in Figure 8, which shows the F$_{H\alpha}$/F$_{UIT}$ for
inter-arm and spiral arm DIG regions in M51. The inter-arm DIG regions
generally have lower ratios than spiral arm DIG (with much overlap),
but this may just be part of a continuous trend with \ha surface
brightness. The faintest DIG regions are the inter-arm regions, but
whether this is a result of the fewer ionizing photons or less gas to
ionize is not clear.

One scenario that could plausibly explain the low
F$_{H\alpha}$/F$_{UIT}$ ratios in the inter-arm DIG is that a lower
fraction of the DIG is ionized by leakage in the inter-arm region. In
a DIG region, the field OB stars emit FUV flux and ionizing photons,
but there is an extra flux of ionizing photons from outside of the
region (leaking into the DIG from an \ion{H}{2} region). Reducing this
extra component of N$_{Lyc}$ would result in a lower
F$_{H\alpha}$/F$_{UIT}$ ratio. If this is the case in the inter-arm
regions, one would expect fainter DIG due to fewer available ionizing
photons. One would also expect the field OB stars to also be more
sparse in the inter-arm region, further reducing the DIG luminosity.

\section{Discussion and Conclusions}

We have compared the FUV and \ha images of 10 nearby spirals in order
to investigate the field star populations and their contribution to
the ionization of the DIG. We find that the F$_{H\alpha}$/F$_{UIT}$
ratio is lower in the DIG than in \ion{H}{2} regions, an indication that
the field OB stars are predominantly later-types. When roughly
corrected for extinction and compared to models of evolving stellar
populations, we find that the DIG F$_{H\alpha}$/F$_{UIT}$ most closely
resemble a steady-state model, or a burst model at an older age than
\ion{H}{2} regions. Comparing the difference between the \ion{H}{2} region and
DIG ratios with that in M33, we find that field OB stars are important
contributors to the DIG ionization in most of the galaxies, and in
some galaxies they may be the dominant ionization source. 

The similarity of the F$_{H\alpha}$/F$_{UIT}$ in the DIG of the sample
of spirals studied here to that of M33 suggests that field OB stars in
these galaxies are responsible for a similar fraction of the
ionization of the DIG as in M33. Evidence for some variation in this
amount is seen, with field stars perhaps dominating the emission in
M51 and NGC 628, but contributing relatively little in NGC 1512 and
NGC 1313 (but still 15\% or more). Most of the galaxies are similar to
M33, however, with field OB star contributions around 40\%. The
variation seen depends on many assumptions, and could be affected by
varying extinction, or it could be a result of differences in the mean
spectral type of the field star population. As a result, the precise
contributions found for each galaxy may not be reliable, and the
strongest result is that field OB stars appear to be an important
ionization source in all of the galaxies. If the contribution is in
fact around 40\% for all spirals, this reduces by almost half the
amount of leakage required from \ion{H}{2} regions to power the DIG. There
are indications that other sources such as shock ionization or
turbulent mixing layers \citep{ssb93} contribute as much as 20\% of
the ionization of the DIG in some galaxies \citep{gwb97,whl97},
further reducing the amount of leakage from \ion{H}{2} regions
required. In fact, \cite{smh00} calculate that cooling supernova
remnants may account for 50\% of the ionization of the DIG in the
Milky Way. The combination of these mechanisms may almost eliminate
the need for density-bounded \ion{H}{2} regions.

Elmegreen (1997, 1998) presented a fractal model of the ISM, in which
ionizing photons can travel at least twice as far as the Str\"omgren
radius. Leakage of ionizing photons from \ion{H}{2} regions is thus
expected in this model, and it should have a large effect on the gas
near \ion{H}{2} regions. It is interesting that we see some indications
that the fraction of DIG ionized by leakage may be lower far from
\ion{H}{2} regions. It seems natural to expect that the DIG near \ion{H}{2}
regions and in spiral arms is ionized by leakage, and the ionization
of the DIG in the inter-arm region has a more important contribution
from field OB stars. This scenario can be tested through spectroscopy
of the DIG, particularly with the $\HeI$ 5876\AA~ and $\OIII$
5007+4959\AA~ lines. These lines both require high energy photons
emitted only by the earliest O stars, so they should be strongest in
DIG ionized by leakage, since early O stars in \ion{H}{2} regions would
contribute to the ionizing spectrum. In the field, very early O stars
are rare \citep{mas95b, hw00}, so these line should be weak. We plan
to carry out this test in the near future.

The fact that field OB star ionization is viable in this sample of 10
spirals suggests that it may be a common feature of spirals. OB stars
outside of \ion{H}{2} regions appear to be fairly important in the energy
balance of the ISM in spirals then, as the results suggest that at
least 15\% of the ionizing photons emitted by OB stars are not emitted
in \ion{H}{2} regions. This fraction may be higher if other sources
contribute to the ionization of the DIG. Since they are in the diffuse
ISM, the influence of these stars may extend much further than their
counterparts in dense \ion{H}{2} regions. By ionizing and heating the
intercloud medium, they may prevent it from condensing, tempering the
formation of molecular clouds, and thus star formation. 

There is clearly a difference between the OB stellar population in the
field and that in \ion{H}{2} regions. The fact that the
F$_{H\alpha}$/F$_{UIT}$ ratios are lower in the DIG in spite of the
addition of ionizing photons through leakage from \ion{H}{2} regions
indicates that later type OB stars dominate in the field. This has
bearing on the origin of this population. Since later-type OB stars
have longer lives than earlier-types, they are more likely to live
long enough to diffuse out of an \ion{H}{2} region, or to survive when an
\ion{H}{2} region is destroyed by SNe from the more massive stars. In this
case the mass function of field OB stars would be truncated at the
stellar mass whose lifetime corresponds to the time it takes OB stars
to become field stars. This time may not be the same for every star
formation event, and since the field OB population is probably a
mixture if stars from several star formation events, the composite IMF
would appear steeper than that for \ion{H}{2} regions. Thus while the mass
function of the field OB population would be steeper than \ion{H}{2}
region stars, the {\it initial} mass function would not. However, if
these stars formed in the field, then the paucity of early-type OB
stars suggests that the {\it initial} mass function is steeper in the
field, and that the process of star formation is fundamentally
different in the field environment. In \cite{hw00} and this paper, no
distinction could be made between field stars that may have drifted
out of \ion{H}{2} regions and those that could only have formed in the
field. The former category certainly exists, but whether it makes up
all or only part of the field OB population is not clear. For LMC
field OB stars, \cite{mas95b} carefully corrected for stars that were
close enough to an \ion{H}{2} region to have drifted out, and the mass
function was still found to be steeper than in \ion{H}{2} regions. Whether
these stars may have originated in \ion{H}{2} regions that no longer exist
is not clear, but is very important in the interpretation of the mass
function slope.

\acknowledgments

We are grateful for the helpful comments from the anonymous
referee. This work benefited from helpful discussions with Salman
Hameed and David Thilker. Bruce Greenawalt obtained and reduced much
of the \ha data. The archival UIT images were obtained through the
Multimission Archive at the Space Telescope Science Institute
(MAST). We would like to acknowledge the UIT project for making their
data available. The availability of the Starburst99 models of
C. Leitherer and collaborators was an immense help in the
interpretation of the observations. This work was supported by grants
from NASA (NAG5--2426) and the NSF (AST 9617014). CH was supported by
a grant from the New Mexico Space Grant Consortium.




{}


\clearpage



\begin{figure}
\figurenum{1}
\epsscale{1.0}
\plotone{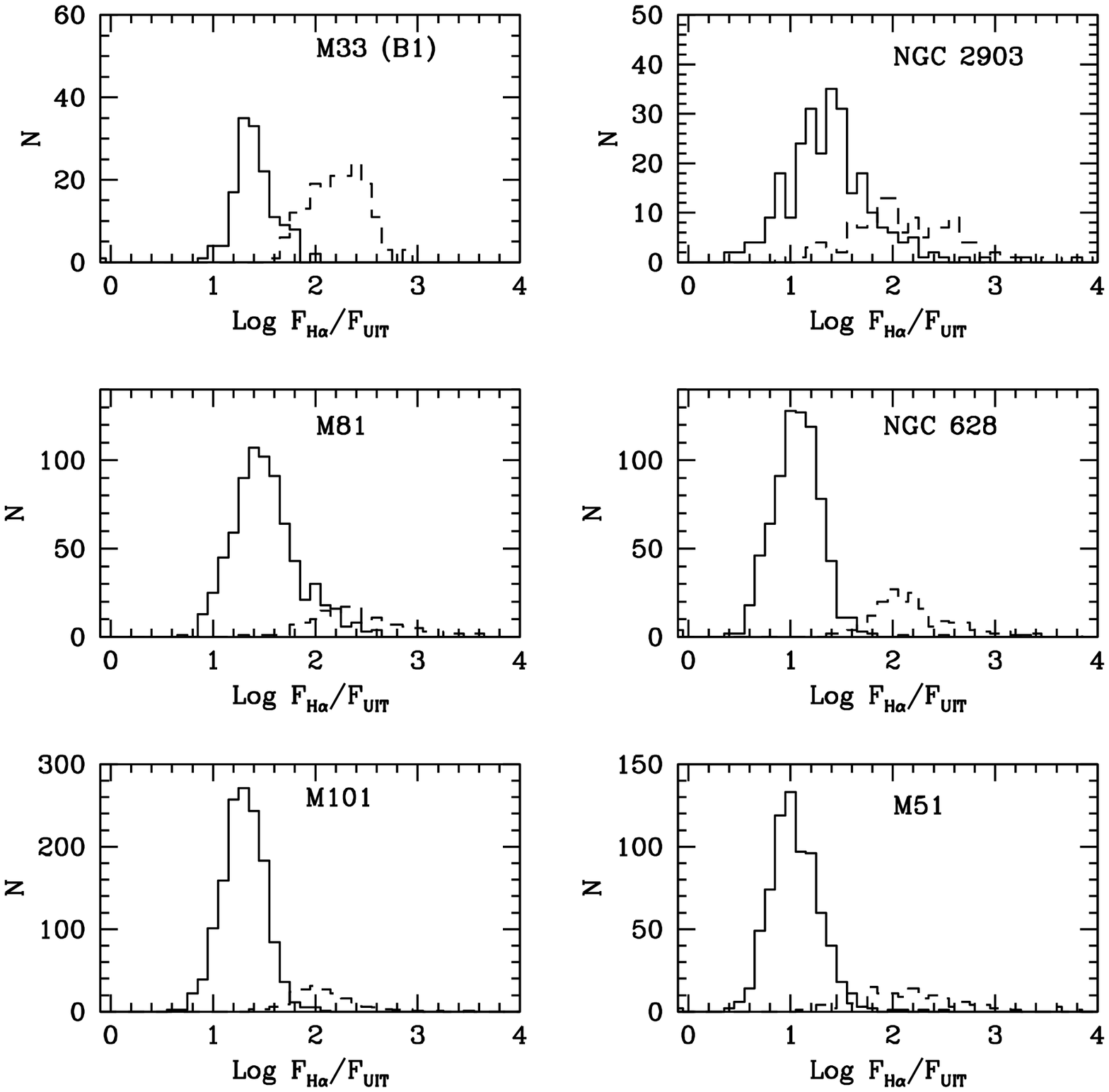}
\caption{Histograms of the
F$_{H\alpha}$/F$_{UIT}$ ratios for the galaxies observed through the
B1 filter. \ion{H}{2} regions are shown with a dashed line, and DIG with a
solid line. These histograms have not been corrected for extinction.}
\end{figure} 

\begin{figure}
\figurenum{2}
\epsscale{1.0}
\plotone{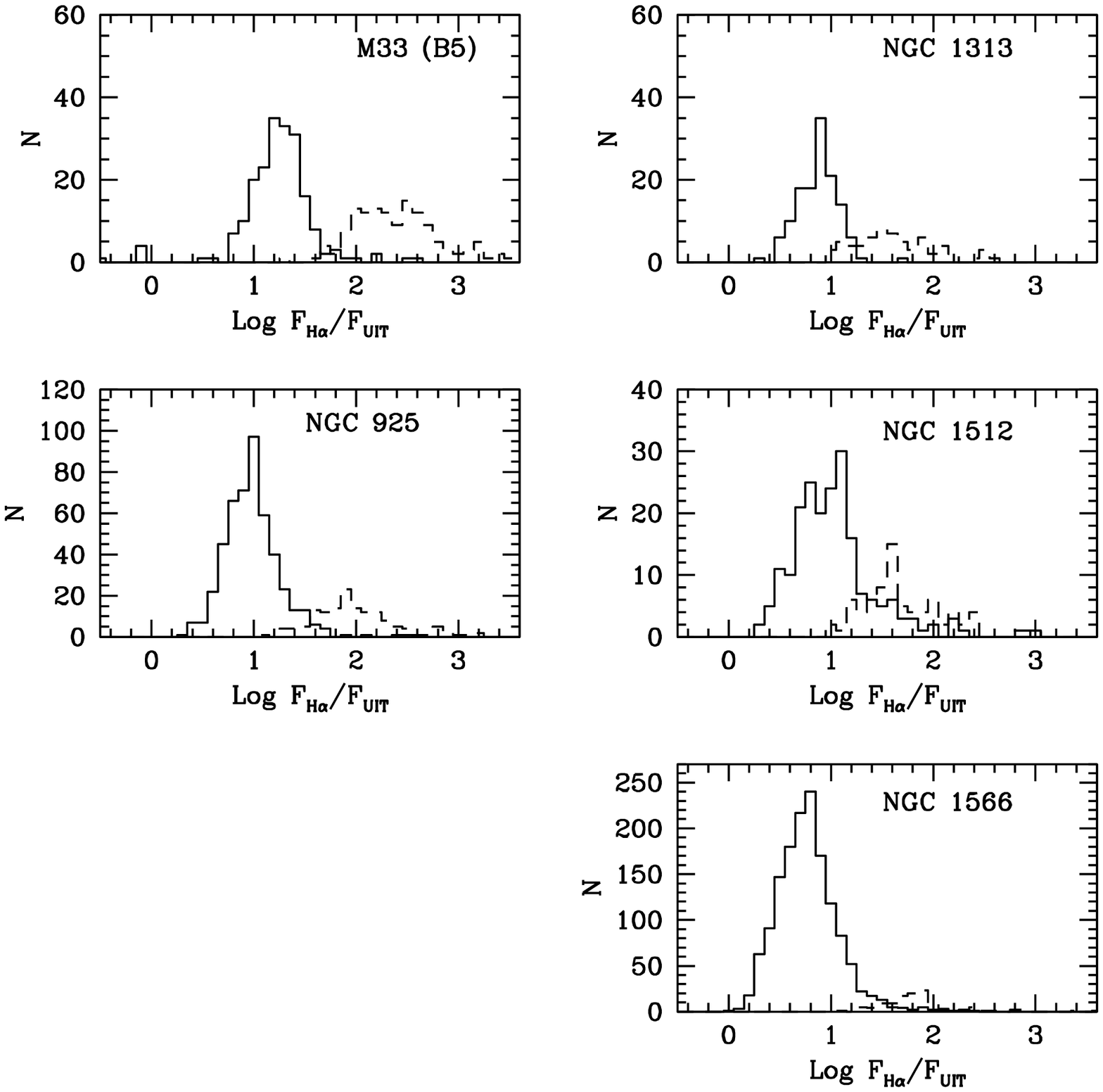}
\caption{Histograms of the
F$_{H\alpha}$/F$_{UIT}$ ratios for the galaxies observed through the
B5 filter. \ion{H}{2} regions are shown with a dashed line, and DIG with a
solid line. These histograms have not been corrected for extinction.}
\end{figure}

\begin{figure}
\figurenum{3}
\epsscale{1.0}
\plotone{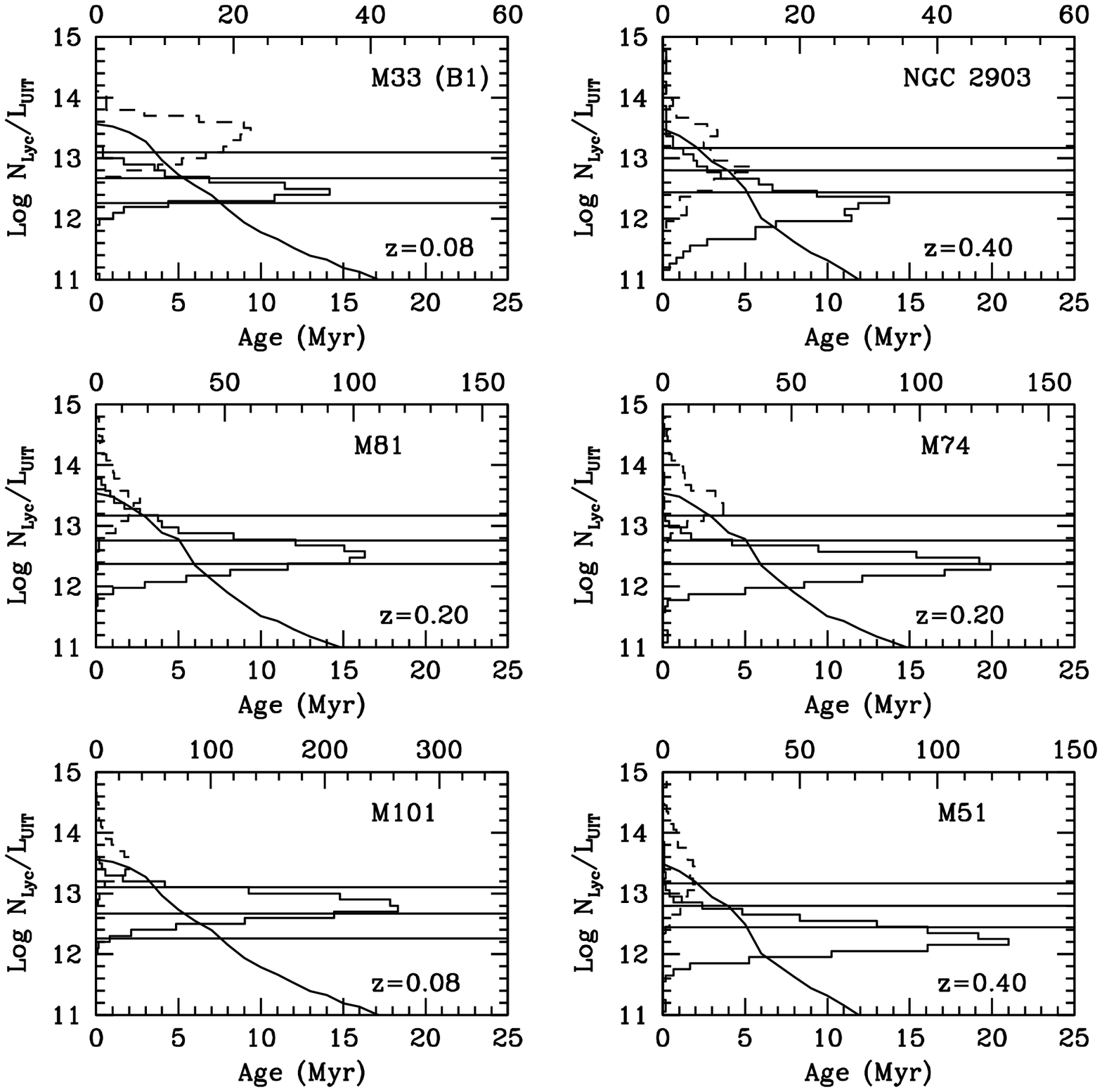}
\caption{Histograms of the $N_{Lyc}/L_{UIT}$ ratios for the galaxies
observed through the B1 filter. \ion{H}{2} regions are shown with a dashed
line, and DIG with a solid line. Starburst99 (Leitherer et al. 1999)
models are also shown. The falling line is a single burst model, and
the horizontal lines are the equilibrium values of steady state models
with constant star formation. The IMF slopes were used in the steady
state models, $\alpha$=2.35 (top line), $\alpha$=3.00 (middle line),
and $\alpha$=3.5 (bottom line). The metallicity of the models for each
galaxy is given in the figure. The histograms have been corrected for
extinction by shifting the \ion{H}{2} regions to match the models, as
described in the text.}
\end{figure} 

\begin{figure}
\figurenum{4}
\epsscale{1.0}
\plotone{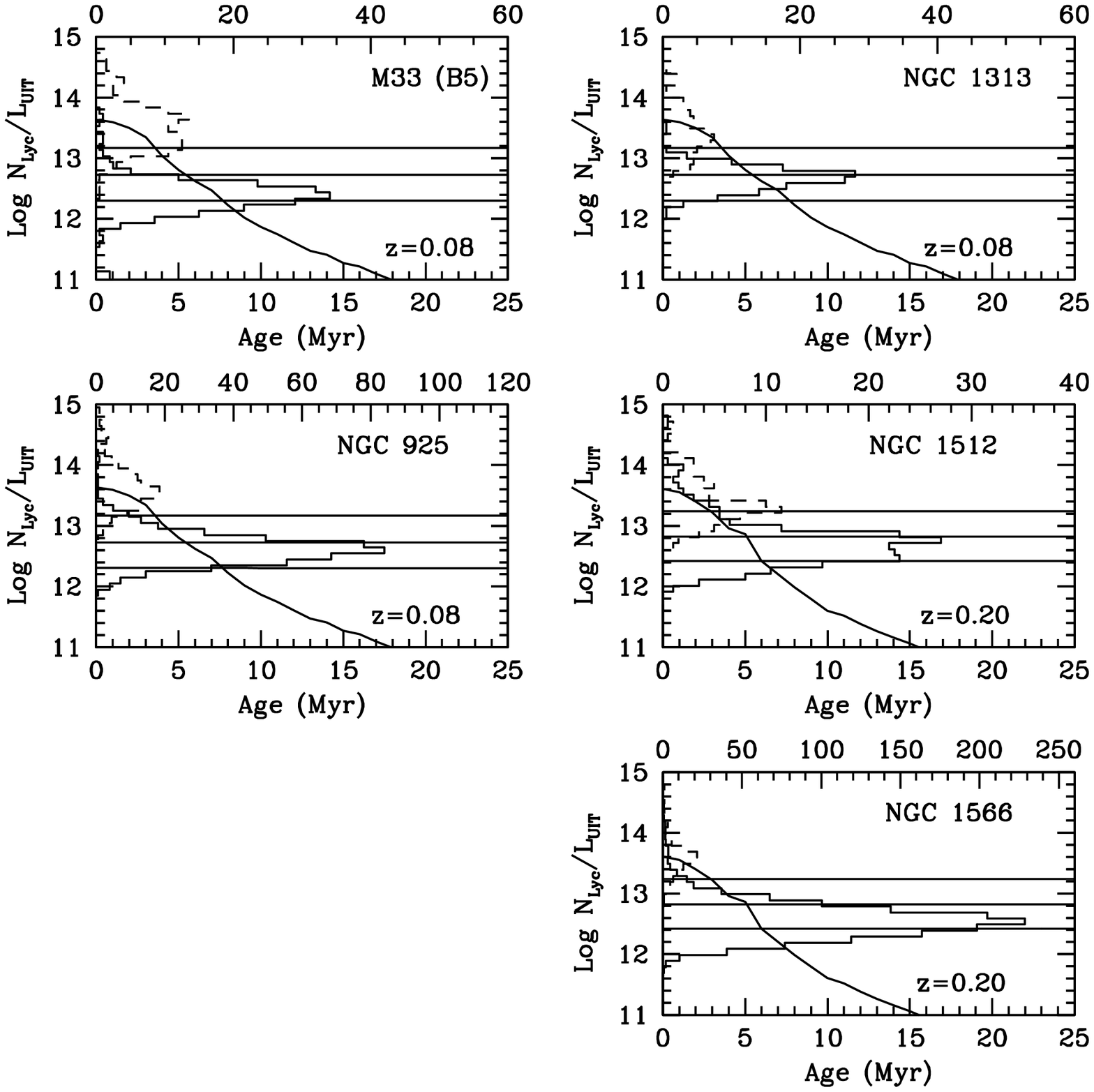}
\caption{Histograms of the $N_{Lyc}/L_{UIT}$
ratios for the galaxies observed through the B5 filter. \ion{H}{2} regions
are shown with a dashed line, and DIG with a solid line. Starburst99
(Leitherer et al. 1999) models are also shown. The falling line is a
single burst model, and the horizontal lines are the equilibrium
values of steady state models with constant star formation. The IMF
slopes were used in the steady state models, $\alpha$=2.35 (top line),
$\alpha$=3.00 (middle line), and $\alpha$=3.5 (bottom line). The
metallicity of the models for each galaxy is given in the figure. The
histograms have been corrected for extinction by shifting the \ion{H}{2}
regions to match the models, as described in the text.}
\end{figure}

\begin{figure}
\figurenum{5}
\epsscale{1.0}
\plotone{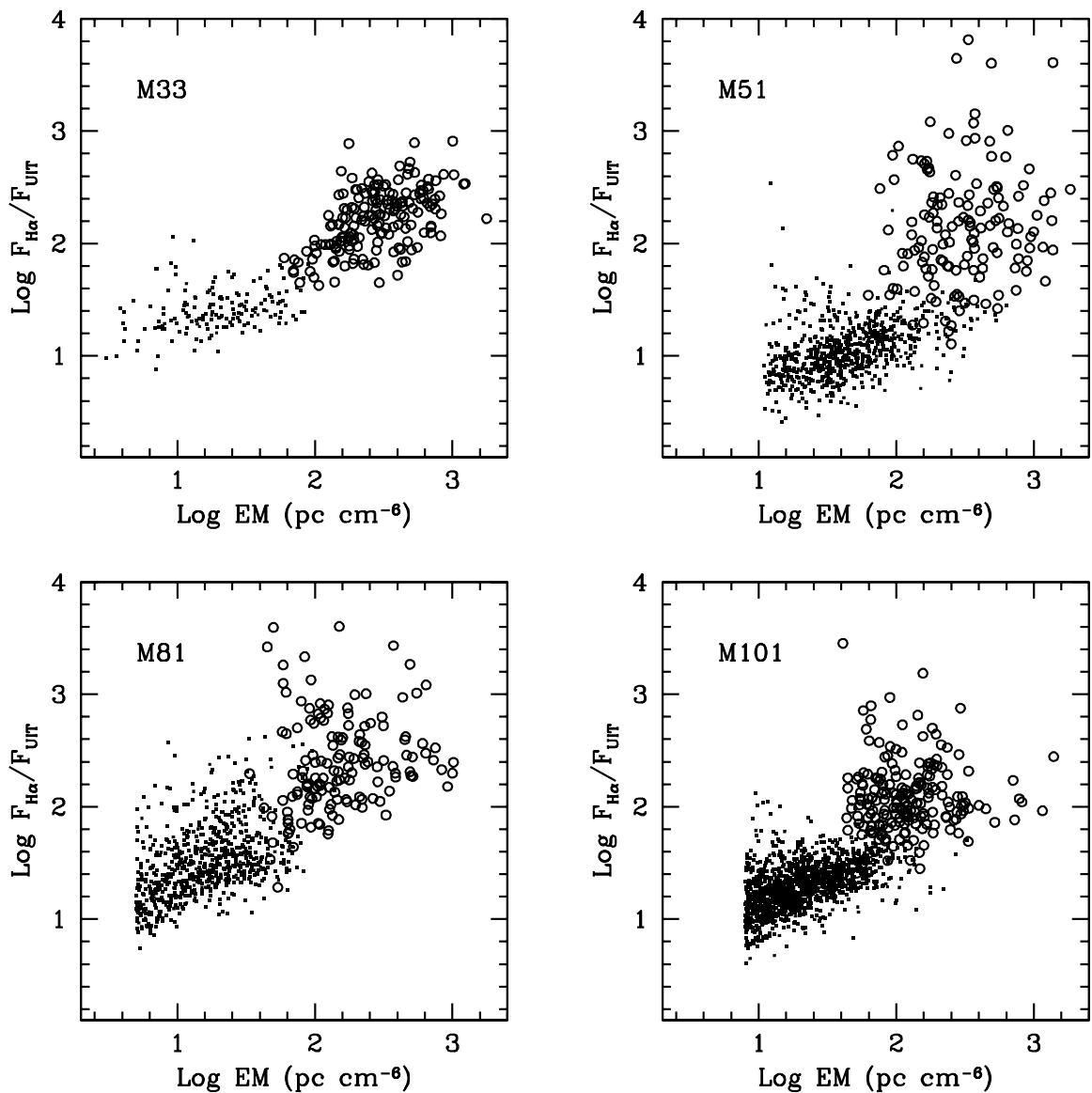}
\caption{The F$_{H\alpha}$/F$_{UIT}$ ratio as a function of \ha surface
brightness. \ion{H}{2} regions are shown as open circles, and the crosses
are DIG regions. The low surface brightness cutoff is defined by the
sensitivity of the \ha image. The ratios were not corrected for extinction.}
\end{figure}

\begin{figure}
\figurenum{6}
\epsscale{1.0}
\plotone{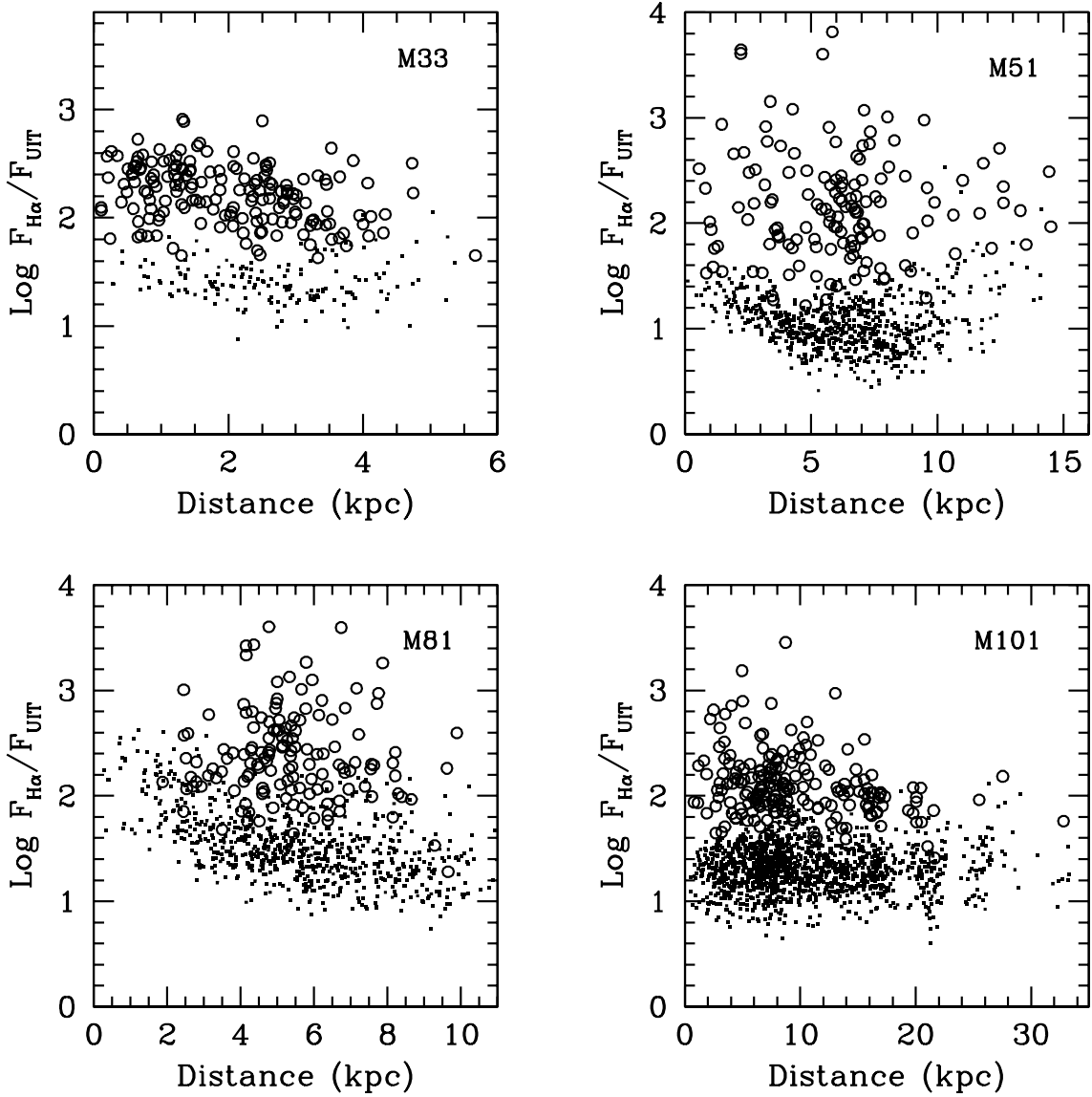}
\caption{The F$_{H\alpha}$/F$_{UIT}$ ratio as a function of projected
galactocentric distance in four galaxies. \ion{H}{2} regions are shown as
open circles, and the points are DIG regions. The ratios were not
corrected for extinction.}
\end{figure}

\begin{figure}
\figurenum{7}
\epsscale{1.0}
\plotone{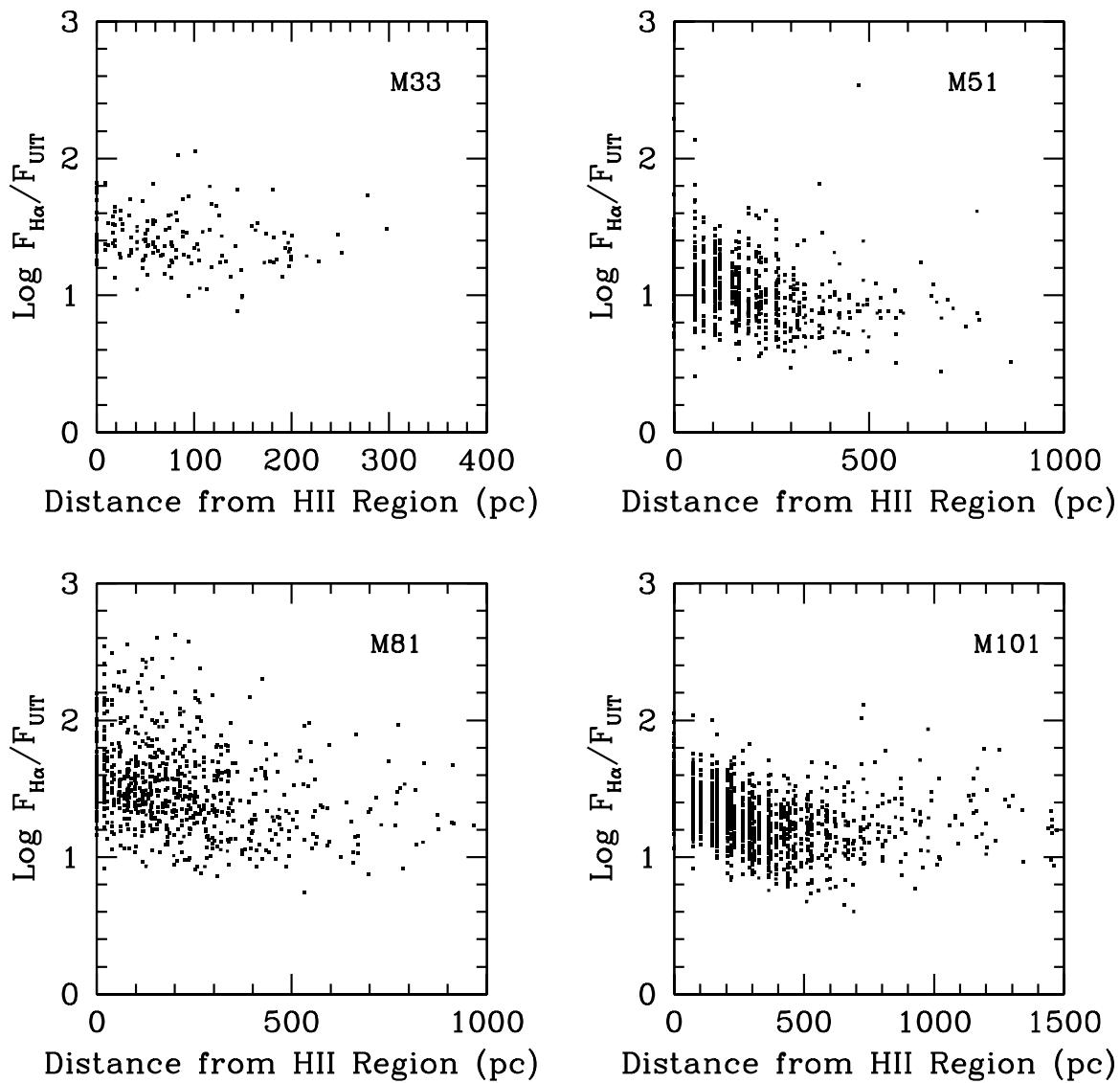}
\caption{The F$_{H\alpha}$/F$_{UIT}$ ratio in the DIG regions as a
function of the projected distance from the nearest \ion{H}{2} region in
four galaxies. Some points have a distance of zero because there is an
\ion{H}{2} region in the DIG box (DIG fluxes were measured on images with
the \ion{H}{2} regions masked out). The ratios were not corrected for
extinction.}
\end{figure} 

\begin{figure}
\figurenum{8}
\epsscale{1.0}
\plotone{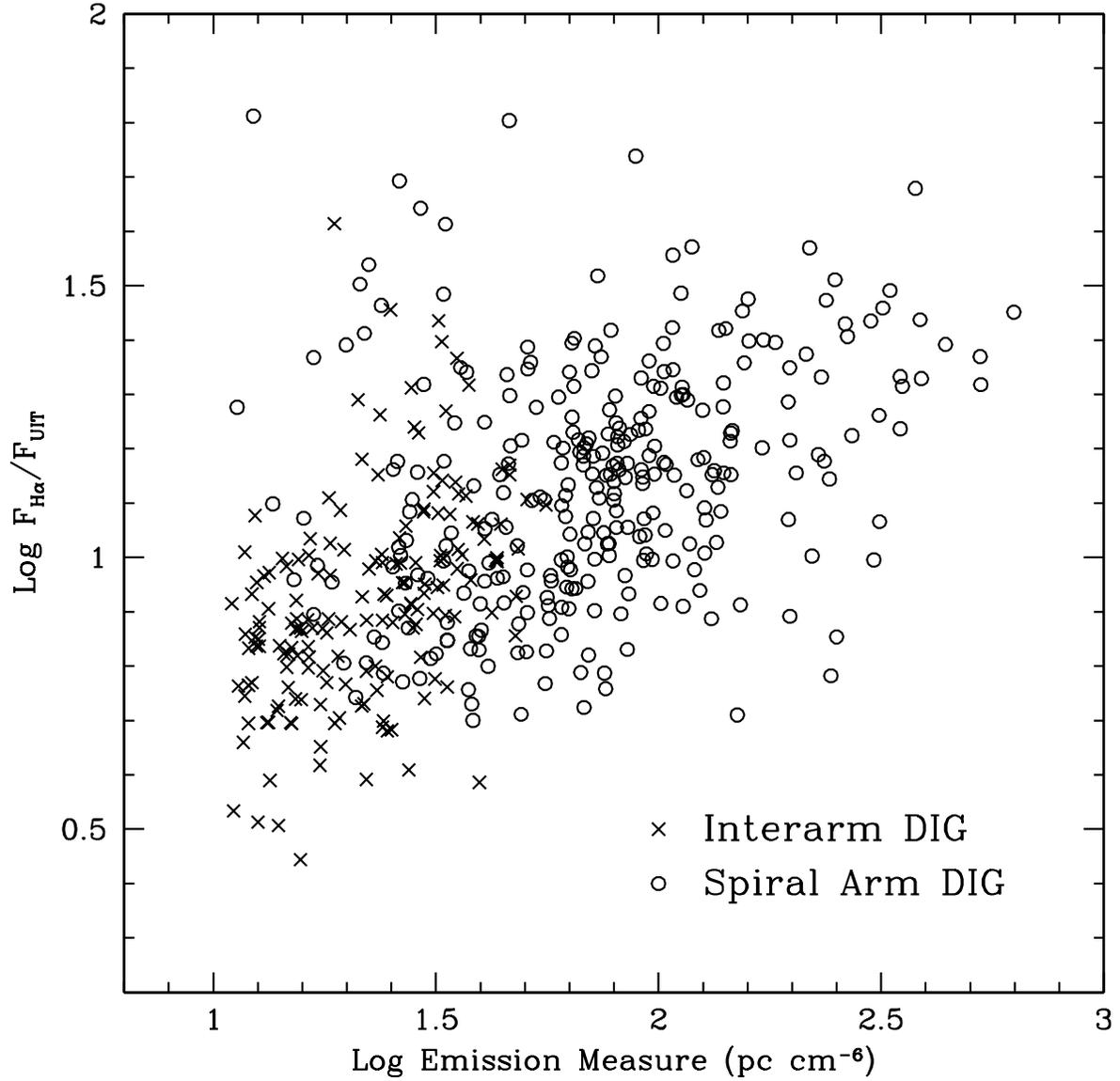}
\caption{The F$_{H\alpha}$/F$_{UIT}$ ratio as a function of \ha surface
brightness for M51 DIG points. The open circles are spiral arm DIG
regions, and the crosses are inter-arm DIG regions. The ratios were not
corrected for extinction.}
\end{figure}

\begin{deluxetable}{lccccc}
\tabletypesize{\scriptsize}
\tablewidth{0pc}
\tablenum{1}
\tablecaption{Galaxy Sample} 
\tablehead{
\colhead{Galaxy}& \colhead{Distance}& \colhead{Hubble Type}& \colhead{12+log(O/H)\tablenotemark{a}} \\ 
\colhead{} & \colhead{(Mpc)}&  \colhead{}& \colhead{}} 

\startdata
NGC 598 (M33)   & 0.84 & Scd & 8.70$\pm$0.06 \\ 
NGC 628 (M74)   & 9.7~  & Sc  & 9.13$\pm$0.16 \\ 
NGC 925         & 9.3~  & Sd  & 8.68$\pm$0.11 \\ 
NGC 1313        & 4.5~  & SBd & 8.33$\pm$0.11 \\ 
NGC 1512        & 16.6\tablenotemark{b} & SBa & \nodata       \\ 
NGC 1566        & 17.4~ & Sbc & 9.10$\pm$0.40 \\ 
NGC 2903        & 6.3~  & Sbc & 9.22$\pm$0.07 \\
NGC 3031 (M81)  & 3.6~  & Sab & 8.89$\pm$0.16 \\ 
NGC 5194 (M51)  & 9.6~  & Sbc & 9.28$\pm$0.11 \\ 
NGC 5457 (M101) & 7.4~  & Scd & 8.76$\pm$0.05 \\
\enddata
\tablenotetext{a}{Oxygen abundance at 3 kpc radius, from Zaritsky et al. 1994. The metallicity of NGC 1512 was not found in the literature.}
\tablenotetext{b}{The distance to NGC 1512 is uncertain. This value was taken from \cite{b88}.}
\end{deluxetable}

\begin{deluxetable}{lcccc}
\tabletypesize{\scriptsize}
\tablewidth{0pc}
\tablenum{2}
\tablecaption{\ha Data} 
\tablehead{
\colhead{Galaxy}&  \colhead{\ha Telescope}&\colhead{\ha Exposure Time}& \colhead{\ha Sensitivity\tablenotemark{a}}& \colhead{\ha 
Calibration} \\ 
\colhead{} &\colhead{} & \colhead{(seconds)}&  \colhead{}&  \colhead{}} 
\startdata
NGC 598 (M33)   & KPNO Schmidt & 20$\times$900 & 0.8 & R-mag       \\ 
NGC 628 (M74)   & KPNO 0.9m    & 16$\times$900 & 0.9 & Standard    \\ 
NGC 925         & KPNO 0.9m    & 16$\times$900 & 1.3 & Standard    \\ 
NGC 1313        & CTIO 0.9m    & 2$\times$900  & 2.2 & Standard    \\ 
NGC 1512        & CTIO 0.9m    & 2$\times$1400 & 1.6 & Standard    \\ 
NGC 1566        & CTIO 0.9m    & 2$\times$1000 & 1.4 & Standard    \\ 
NGC 2903        & KPNO 0.9m    & 16$\times$900 & 0.7 & Standard    \\
NGC 3031 (M81)  & KPNO Schmidt & 14$\times$900 & 1.5 & $\HII$ Regions \\ 
NGC 5194 (M51)  & KPNO 0.9m    & 3$\times$1200 & 2.7 & R-mag       \\ 
NGC 5457 (M101) & KPNO Schmidt & 13$\times$900 & 1.6 & $\HII$ Regions \\
\enddata
\tablenotetext{a}{$\times$10$^{17}$erg s$^{-1}$ cm$^{-2}$ arcsec$^{-2}$}
\end{deluxetable}

\begin{deluxetable}{lccccc}
\tabletypesize{\scriptsize}
\tablewidth{0pc}
\tablenum{3}
\tablecaption{UIT Data} 
\tablehead{
\colhead{Galaxy}& \colhead{Exposure Time}&
\colhead{Filter\tablenotemark{a}}&
\colhead{Sensitivity\tablenotemark{b}}& \colhead{Resolution\tablenotemark{c}} & \colhead{Mission} \\
\colhead{} & \colhead{(seconds)}& \colhead{}& \colhead{} & \colhead{(parsecs)} & \colhead{}} 
\startdata
NGC 598 (M33)   & 424  & B1 & 1.3 & 8 & Astro-1 \\ 
NGC 598 (M33)   & 626  & B5 & 2.0 & 8 & Astro-2 \\ 
NGC 628 (M74)   & 514  & B1 & 0.9 & 70 & Astro-1 \\ 
NGC 925         & 1590 & B5 & 0.7 & 90 & Astro-2 \\ 
NGC 1313        & 1070 & B5 & 1.0 & 44 & Astro-2 \\ 
NGC 1512        & 948  & B5 & 1.1 & 161 & Astro-2 \\ 
NGC 1566        & 1390 & B5 & 0.8 & 169 & Astro-2 \\ 
NGC 2903        & 548  & B1 & 1.0 & 61 & Astro-2 \\
NGC 3031 (M81)  & 640  & B1 & 0.8 & 35 & Astro-1 \\ 
NGC 5194 (M51)  & 1100 & B1 & 0.5 & 93 & Astro-2 \\ 
NGC 5457 (M101) & 1310 & B1 & 0.5 & 72 & Astro-2 \\
\enddata
\tablenotetext{a}{The UIT B1 filter is 354\AA~ wide with an effective wavelength of $\lambda_{eff}$=1521\AA, and the B5 filter is
 225\AA~ wide with $\lambda_{eff}$=1615\AA.}
\tablenotetext{b}{$\times$10$^{17}$erg s$^{-1}$ cm$^{-2}$ arcsec$^{-2}$}
\tablenotetext{c}{Spatial resolution based on 2$^{\prime\prime}$ angular resolution of UIT.}
\end{deluxetable}

\begin{deluxetable}{lcccccc}
\tabletypesize{\scriptsize}
\tablewidth{0pc}
\tablenum{4}
\tablecaption{Measured Properties} 
\tablehead{
\colhead{Galaxy} & \colhead{L$_{H\alpha}$\tablenotemark{a}}& \colhead{\ha
DF\tablenotemark{a}} & \colhead{log L$_{H\alpha}$/L$_{UIT}$ Total} &
\colhead{log L$_{H\alpha}$/L$_{UIT}$ $\HII$} &
\colhead{log L$_{H\alpha}$/L$_{UIT}$ DIG} & FUV DF\tablenotemark{b} \\
\colhead{} & \colhead{(erg s$^{-1}$)}& \colhead{}& \colhead{}& \colhead{}& \colhead{}& \colhead{} } 
\startdata
NGC 598 (B1)    & 3.0$\pm$0.3$\times$10$^{40}$ & 0.38$\pm$0.03 & 1.75$\pm$0.10 & 2.06$\pm$0.07 & 1.47$\pm$0.10 & 0.73$\pm$0.21 \\ 
NGC 628         & 1.7$\pm$0.3$\times$10$^{41}$ & 0.43$\pm$0.05 & 1.64$\pm$0.11 & 2.41$\pm$0.09 & 1.31$\pm$0.11 & 0.91$\pm$0.26 \\ 
NGC 2903        & 6.5$\pm$1.8$\times$10$^{40}$ & 0.27$\pm$0.09 & 2.16$\pm$0.14 & 2.54$\pm$0.12 & 1.74$\pm$0.14 & 0.72$\pm$0.21 \\
NGC 3031        & 7.8$\pm$1.2$\times$10$^{40}$ & 0.49$\pm$0.08 & 2.12$\pm$0.11 & 2.65$\pm$0.08 & 1.87$\pm$0.11 & 0.87$\pm$0.25 \\ 
NGC 5194        & 3.1$\pm$0.3$\times$10$^{41}$ & 0.43$\pm$0.03 & 1.88$\pm$0.10 & 2.22$\pm$0.07 & 1.62$\pm$0.10 & 0.77$\pm$0.22 \\ 
NGC 5457        & 2.4$\pm$0.3$\times$10$^{41}$ & 0.43$\pm$0.04 & 1.41$\pm$0.11 & 1.95$\pm$0.08 & 1.11$\pm$0.11 & 0.84$\pm$0.24 \\
NGC 598 (B5)    & 3.0$\pm$0.3$\times$10$^{40}$ & 0.38$\pm$0.03 & 1.51$\pm$0.10 & 1.99$\pm$0.07 & 1.17$\pm$0.10 & 0.84$\pm$0.24 \\ 
NGC 925         & 7.6$\pm$0.5$\times$10$^{40}$ & 0.37$\pm$0.05 & 1.70$\pm$0.10 & 2.30$\pm$0.06 & 1.33$\pm$0.10 & 0.86$\pm$0.25 \\ 
NGC 1313        & 2.7$\pm$0.5$\times$10$^{40}$ & 0.46$\pm$0.09 & 1.57$\pm$0.12 & 1.88$\pm$0.09 & 1.35$\pm$0.12 & 0.76$\pm$0.22 \\ 
NGC 1512        & 2.9$\pm$0.6$\times$10$^{40}$ & 0.15$\pm$0.03 & 1.25$\pm$0.12 & 2.20$\pm$0.10 & 0.47$\pm$0.12 & 0.91$\pm$0.26 \\ 
NGC 1566        & 3.2$\pm$0.6$\times$10$^{41}$ & 0.33$\pm$0.07 & 1.58$\pm$0.12 & 2.13$\pm$0.09 & 1.18$\pm$0.12 & 0.82$\pm$0.23 \\ 
\enddata
\tablenotetext{a}{The \ha filter used for NGC 3031, NGC 5194, and NGC
5457 contains a contribution from $\NII$.}
\tablenotetext{b}{The diffuse fraction (DF) is the ratio of the
luminosity contributed by DIG to the total luminosity of the galaxy.}
\end{deluxetable}

\begin{deluxetable}{lccccc}
\tabletypesize{\scriptsize}
\tablewidth{0pc}
\tablenum{5}
\tablecaption{Extinction Corrections} 
\tablehead{
\colhead{Galaxy}& \colhead{Mean Observed Log $N_{Lyc}/L_{UIT}$} & \colhead{Difference\tablenotemark{a}} & \colhead{E(B-V) MW\tablenotemark{b}} & \colhead{E(B-V) LMC\tablenotemark{b}} & \colhead{Published E(B-V)\tablenotemark{c}}}
\startdata
NGC 598 (B1)    & 14.27 & 0.75 & 0.33 & 0.23 & $0.05 - 0.33$\\ 
NGC 628         & 14.05 & 0.57 & 0.25 & 0.18 & $0.47 - 0.52$\\ 
NGC 2903        & 14.35 & 0.98 & 0.44 & 0.31 & $0.68 - 1.34$\\
NGC 3031        & 14.25 & 0.77 & 0.34 & 0.24 & $0.14 - 0.48$\\ 
NGC 5194        & 13.96 & 0.59 & 0.27 & 0.19 & $0.48 - 1.62$\\ 
NGC 5457        & 13.87 & 0.35 & 0.15 & 0.11 & $0.00 - 0.40$\\
NGC 598 (B5)    & 14.31 & 0.71 & 0.33 & 0.23 & $0.05 - 0.33$\\ 
NGC 925         & 13.79 & 0.20 & 0.09 & 0.07 & $0.16 - 0.48$\\ 
NGC 1313        & 13.65 & 0.05 & 0.02 & 0.02 & $0.31 - 0.83$\\ 
NGC 1512        & 13.68 & 0.13 & 0.06 & 0.04 & $0.18 - 1.06$\\ 
NGC 1566        & 13.61 & 0.06 & 0.03 & 0.02 & $0.55 - 0.74$\\ 
\enddata
\tablenotetext{a}{The difference between the observed Log
$N_{Lyc}/L_{UIT}$ and that predicted for a 1 Myr burst.}
\tablenotetext{b}{The amount of extinction which would account for the
difference. The Milky Way extinction law is
from Cardelli et al. 1989,
and the LMC extinction law is from Howarth 1983.}
\tablenotetext{c}{References for published reddening: NGC~598 -- Massey et al. 1995a; NGC~628 -- Petersen \& Gammelgaard 1996; NGC~2903, NGC~1512, NGC~5194 -- Quillen \& Yukita 2001; NGC~3031 -- Kaufman et al. 1987; NGC~5457 -- Bresolin, Kennicutt, \& Stetson 1996; NGC~925 -- Zaritsky et al. 1994; NGC~1313 -- Walsh \& Roy 1997; NGC~1566 -- Roy \& Walsh 1986 (c(H$\beta$) was converted to E(B-V) for NGC~1566).  }
\end{deluxetable}

\begin{deluxetable}{lccc}
\tabletypesize{\scriptsize}
\tablewidth{0pc}
\tablenum{6}
\tablecaption{Galaxy Comparison} 
\tablehead{
\colhead{Galaxy} & 
\colhead{Log (mean F$_{H\alpha}$/F$_{UIT}$) HII $-$ Log (mean F$_{H\alpha}$/F$_{UIT}$) DIG} &
\colhead{Fraction of M33 Ratio} & 
\colhead{Field Star Contribution (\%)}}
\startdata
NGC 598 (B1)    & 0.81$\pm$0.10 & \nodata        & 0.40$\pm$0.12\tablenotemark{a}  \\ 
NGC 628         & 1.17$\pm$0.11 & 0.44$\pm$0.17  & 0.91$\pm$0.17  \\ 
NGC 2903        & 0.73$\pm$0.14 & 1.20$\pm$0.55  & 0.33$\pm$0.19  \\
NGC 3031        & 0.87$\pm$0.11 & 0.87$\pm$0.34  & 0.46$\pm$0.21  \\ 
NGC 5194        & 1.09$\pm$0.10 & 0.52$\pm$0.20  & 0.77$\pm$0.22  \\ 
NGC 5457        & 0.78$\pm$0.11 & 1.07$\pm$0.41  & 0.37$\pm$0.17  \\
NGC 598 (B5)    & 1.16$\pm$0.10 & \nodata        & 0.40$\pm$0.12\tablenotemark{a}  \\ 
NGC 925         & 1.02$\pm$0.10 & 1.38$\pm$0.52  & 0.29$\pm$0.13  \\ 
NGC 1313        & 0.81$\pm$0.12 & 2.24$\pm$0.92  & 0.18$\pm$0.09  \\ 
NGC 1512        & 0.72$\pm$0.12 & 2.75$\pm$1.15  & 0.15$\pm$0.08  \\ 
NGC 1566        & 1.13$\pm$0.12 & 1.07$\pm$0.44  & 0.37$\pm$0.18  \\ 
\enddata
\tablenotetext{a}{The field OB star contribution given for M33 was that measured in Hoopes \& Walterbos 2000.}
\end{deluxetable}


\begin{thebibliography}{}

\bibitem[Bell \& Kennicutt(2001)]{bk01} Bell, E. F. \& Kennicutt, R. C. 2001, ApJ, 548, 681 

\bibitem[Bresolin, Kennicutt, \& Stetson(1996)]{bks96} Bresolin, F., Kennicutt, R.\ C., \& Stetson, P.\ B.\ 1996, \aj, 112, 1009 


\bibitem[Buta(1988)]{b88}Buta, R.\ 1988, ApJS, 66, 233 

\bibitem[Cardelli, Clayton, \& Mathis(1989)]{ccm89}Cardelli, J. A., Clayton, G. C., \& Mathis, J. S. 1989, ApJ, 345, 245

\bibitem[Cornett et al.(1994)]{c94} Cornett, R. H. et al. 1994, ApJ, 426, 553 

\bibitem[Elmegreen(1997)]{e97}Elmegreen, B. G. 1997, ApJ, 477, 196

\bibitem[Elmegreen(1998)]{e98}Elmegreen, B. G. 1998, PASA, 15, 74

\bibitem[Ferguson et al.(1996)]{f96}Ferguson, A. M. N., Wyse, R. F. G., Gallagher, J. S., Hunter, D. A. 1996, AJ, 111, 2265

\bibitem[Galarza, Walterbos, \& Braun(1999)]{gwb99}Galarza, V. C., Walterbos, R. A. M., \& Braun, R. 1999, AJ, 118, 2775

\bibitem[Greenawalt(1998)]{g98}Greenawalt, B. 1998, Ph.D. Thesis, New Mexico State University

\bibitem[Greenawalt, Walterbos, \& Braun(1997)]{gwb97}Greenawalt, B., Walterbos, R. A. M., \& Braun, R. 1997, ApJ, 483, 666

\bibitem[Greenawalt et al.(1998)]{gwth98}Greenawalt, B., Walterbos, R. A. M., Thilker, D. A., \& Hoopes, C. G. 1998, ApJ, 506, 135

\bibitem[Haffner, Reynolds, \& Tufte(1999)]{hrt99}Haffner, L. M., Reynolds, R. J., \& Tufte, S. L. 1999, ApJ, 523, 223

\bibitem[Heiles et al.(1995)]{hklr96} Heiles, C., Koo, B., Levenson, N. A. \& Reach, W. T. 1996, ApJ, 462, 326 

\bibitem[Hill et al.(1995)]{hill95}Hill, J. K. et al. 1995, ApJ, 438, 181

\bibitem[Hill et al.(1998)]{hill98}Hill, R. S. et al. 1998, ApJ, 507, 179

\bibitem[Hoopes \& Walterbos(2000)]{hw00}Hoopes, C. G. \& Walterbos, R. A. M. 2000, ApJ, 541, 597

\bibitem[Hoopes, Walterbos, \& Greenawalt(1996)]{hwg96}Hoopes, C. G., Walterbos, R. A. M., \& Greenawalt, B. 1996, AJ, 112, 1429

\bibitem[Howarth(1983)]{h83}Howarth, I. D. 1983, MNRAS, 203, 301

\bibitem[Jones, Tielens \& Hollenbach(1996)]{jth96} Jones, A. P., Tielens, A. G. G. M. \& Hollenbach, D. J. 1996, ApJ, 469, 740 

\bibitem[Kaufman et al.(1987)]{K87} Kaufman, M., Bash, F.\ N., Kennicutt, R.\ C., \& Hodge, P.\ W.\ 1987, \apj, 319, 61 


\bibitem[Leitherer et al.(1999)]{l99}Leitherer, C., Schaerer, D. Goldader, J. D., Gonz\'alez Delgado, R. M., Robert, C., Foo Kune, D., de Mello, D. F., Devost, D., \& Heckman, T. M. 1999, ApJS, 123, 3

\bibitem[Massey(1985)]{mas85}Massey, P. 1985, PASP, 97, 5

\bibitem[Massey(1998)]{mas98}Massey, P. 1998, in ASP Conf. Proc. 142, The Stellar Initial Mass Function, ed. G. Gilmore \& D. Howell (San Francisco:ASP), 17

\bibitem[Massey et al.(1995a)]{mas95a}Massey, P., Armandroff, T. E., Pyke, R., Patel, K., \& Wilson, C. D. 1995a, AJ, 110, 2715

\bibitem[Massey et al.(1995b)]{mas95b}Massey, P., Lang, C. C., DeGioia-Eastwood, K., \& Garmany, C. D. 1995b, ApJ, 438, 188

\bibitem[McKee \& Williams(1997)]{mw97}McKee, C. F. \& Williams, J. P. 1997, ApJ, 476, 144

\bibitem[Oey \& Kennicutt(1997)]{ok97}Oey, M. S. \& Kennicutt, R. C. 1997, MNRAS, 291, 827

\bibitem[Oey \& Clarke(1998)]{oc98} Oey, M. S. \& Clarke, C. J. 1998, AJ, 115, 1543 


\bibitem[Osterbrock(1989)]{o89} Osterbrock, D. E. 1989, Astrophysics of Gaseous Nebulae and Active Galactic Nuclei (Mill Valley: University Science Books)

\bibitem[Patel \& Wilson(1995)]{pw95}Patel, K., \& Wilson, C. D. 1995, ApJ, 451, 607

\bibitem[Petersen \& Gammelgaard(1996)]{pg96} Petersen, L.\ \& Gammelgaard, P.\ 1996, \aap, 308, 49 

\bibitem[Rand(1997)]{ran97} Rand, R. J. 1997, ApJ, 474, 129

\bibitem[Quillen \& Yukita(2001)]{qy01} Quillen, A.\ C.\ \& Yukita, M.\ 2001, \aj, 121, 2095 



\bibitem[Reynolds(1991)]{re91} Reynolds, R. J. 1991, in The Interstellar Disk-Halo Connection in Galaxies, IAU Symposium No. 144, edited by H. Bloemen (Dordrecht:Kluwer), 67

\bibitem[Reynolds \& Tufte(1995)]{rt95} Reynolds, R. J. \& Tufte, S. L. 1995, ApJ, 439, L17

\bibitem[Roy \& Walsh(1986)]{rw86} Roy, J.\ \& Walsh, J.\ R.\ 1986, \mnras, 223, 39 

\bibitem[Schaerer \& de Koter(1997)]{sd97} Schaerer, D.\ \& 
de Koter, A.\ 1997, \aap, 322, 598 

\bibitem[Schaerer, et al.(1993)]{s93} 
Schaerer, D., Meynet, G., Maeder, A., \& Schaller, G.\ 1993, \aaps, 98, 523 

\bibitem[Slavin, McKee, \& Holenbach(2000)]{smh00}Slavin, J. D., McKee, C. F., \& Hollenbach, D. J. 2000, ApJ, 541, 218

\bibitem[Slavin, Shull, \& Begelman(1993)]{ssb93}Slavin, J. D., Shull, J. M., \& Begelman, M. C. 1993, ApJ, 407, 83

\bibitem[Stecher et al.(1997)]{s97}Stecher, T., et al. 1997, PASP, 109, 584

\bibitem[Stewart et al.(2000)]{s00}Stewart, S. G., et al. 2000, ApJ, 529, 201

\bibitem[Torres-Peimbert, Lazcano-Araujo, \& Peimbert(1974)]{tlp74}Torres-Peimbert, S., Lazcano-Araujo, A., \& Peimbert, M. 1974, ApJ, 191, 401

\bibitem[von Hippel \& Bothun(1990)]{vb90}von Hippel, T., \& Bothun, G. 1990, AJ, 100, 403

\bibitem[Walterbos \& Braun(1994)]{wb94}Walterbos, R. A. M., \& Braun, R. 1994, ApJ, 431, 156

\bibitem[Walterbos \& Braun(1996)]{wb96}Walterbos, R. A. M., \& Braun, R. 1996, in ASP Conf. Proc. 106, The Minnesota Lectures on Extragalactic Neutral Hydrogen, ed. E. D. Skillman (San Francisco:ASP), 1

\bibitem[Walsh \& Roy(1997)]{wr97} Walsh, J.\ R.\ \& Roy, J.\ -.\ 1997, \mnras, 288, 726 

\bibitem[Wang et al.(1997)]{whl97}Wang, J., Heckman, T. M., \& Lehnert, M. D. 1997, ApJ, 491, 114

\bibitem[Zaritsky et al.(1994)]{zkh94}Zaritsky, D., Kennicutt, R. C., \& Huchra, J. P. 1994, ApJ, 420, 87


\end{thebibliography}
\end{document}